\documentclass[aps,pra,twocolumn,superscriptaddress,notitlepage]{revtex4-2}
\usepackage{CJK}
\usepackage[pdftex]{graphicx}
\usepackage{float}
\usepackage{amsmath}
\usepackage{amssymb}
\usepackage{url}
\usepackage{natbib}

\usepackage[colorlinks,
                   linkcolor=blue,
                   anchorcolor=blue,
                   citecolor=blue,
                   urlcolor=blue]
                  {hyperref}
\begin{document}
\begin{CJK*}{UTF8}{gbsn}
\title{Observation of Time Crystal in a Spin Maser System}
\author{Weiyu Wang}
\author{Mingjun Feng}
\author{Qianjin Ma}
\affiliation{National Time Service Center, Chinese Academy of Sciences, Xi'an 710600, China}
\affiliation{Key Laboratory of Time Reference and Applications, Chinese Academy of Sciences, Xi’an 710600, China}
\affiliation{University of Chinese Academy of Sciences, Beijing 100049, China}
\author{Zi Cai}

\affiliation{Wilczek Quantum Center and Key Laboratory of Artificial Structures and Quantum Control, School of Physics and Astronomy, Shanghai Jiao Tong University, Shanghai 200240, China}
\affiliation{Shanghai Research Center for Quantum Sciences, Shanghai 201315, China}
\author{Erwei Li}
\author{Guobin Liu}
\email{Corresponding author: liuguobin@ntsc.ac.cn}

\affiliation{National Time Service Center, Chinese Academy of Sciences, Xi'an 710600, China}
\affiliation{Key Laboratory of Time Reference and Applications, Chinese Academy of Sciences, Xi’an 710600, China}
\affiliation{University of Chinese Academy of Sciences, Beijing 100049, China}

\date{\today}
\begin{abstract}
Pair interaction potentials between atoms in a crystal are in general non-monotonic in distance, with a local minimum whose position gives the lattice constant of the crystal.  A temporal analogue of this idea of crystal formation is still pending despite intensive studies on the time crystal phase.  In a hybrid spin maser system with a time delay feedback, we report the observation of a time crystal induced by a retarded interaction with a characteristic time scale.  This nonequilibrium phase features  a self-sustained oscillation with an emergent frequency other than the intrinsic Larmor precession frequency of the spin maser system. It is shown that the amplitude of the oscillation is robust against perturbation, while its time phase randomly distributes from 0 to $2\pi$ for different realizations, a signature of spontaneous time translation symmetry breaking. This time crystal phase emerges only when the feedback strength exceeds a critical value, at which the system experiences a first order phase transition. Such a retarded interaction induced time crystal is closer to the idea of crystal, compared to other time crystal realizations.
\end{abstract}
\maketitle
\end{CJK*}

\section{Introduction}
Crystals in nature  are composed of atoms with self-organized periodic structures, which spontaneously break the continuous spatial translation symmetry into discrete ones. This idea was generalized into temporal domain\cite{Wilczek2012}, giving rise to an intriguing phase dubbed ``time crystal''\cite{Sacha2018,Else2020,Yao2023} that has attracted considerable interests in the past decade\cite{Bruno2013,Watanabe2015,Sacha2015,Else2016,Khemani2016,Yao2017,Choi2017,Zhang2017,Cai2020,Trager2021,Yang2021,Stehouwer2021,Kyprianidis2021,Mi2022,Frey2022,Randall2021}. Time crystals can be classified as discrete or continuous depending on whether the broken translation symmetry is discrete or continuous. The time crystal has been observed in various systems including atom-cavity\cite{Kongkhambut2022}, semiconductor\cite{Greilich2024}, exciton-polariton\cite{Haddad2024} and thermal Rydberg gas systems\cite{Wu2024}, while the underlying mechanism behind most time crystals to date is the limit cycle: an asymptotic periodic solution of nonlinear differential equations corresponding to a closed phase space trajectory robust against perturbations. Searching for time crystal beyond this limit cycle scenario remains challenging for both theorists and experimentalists. A kind of time quasicrystal was recently demonstrated in the superfluid $^3$He system, which forms an interesting new platform studying time crystal dynamics \cite{Autti2018,Autti2021,Autti2022}.

The interaction potential in a crystal is in general non-monotonic in distance,  with a potential minimum whose position provides a key ingredient of a crystal: the lattice constant. A temporal analogue of this picture  requires a retarded interaction which is nonlocal in time and with a characteristic time scale akin to the lattice constant. Compared to the widely existed nonlocal interactions in space, time-delay interactions are much rarer in nature: they usually appear as effective interactions induced by natural environments or artificial feedback protocols. Feedback is procedure of modifying  system parameters according to the measurement outcomes, which plays an important role in physics and engineering\cite{Magann2022,Yamaguchi2023,Wu2022,Terhal2015,Hurst2020,Lloyd2000,Munoz2020,McGinley2022,Ivanov2020,Wu2023}. In most realistic systems, feedback is not instantaneous but  accompanied by a time delay with a characteristic time scale, thus can be considered as a source of retarded interaction. Take a spin system for an example, if the magnetization of spins are continuously measured, and fed back into the system Hamiltonian with a time delay $\tau$, thus the dynamics of the spin at time $t$  depends on  the spin magnetization in earlier time ($t-\tau$). As a consequence, such a feedback procedure actually builds up an effective retarded interaction with a characteristic time scale ($\tau$)  between spins at different times.

\begin{figure}[htbp]
\centering
\includegraphics[width=0.4\textwidth]{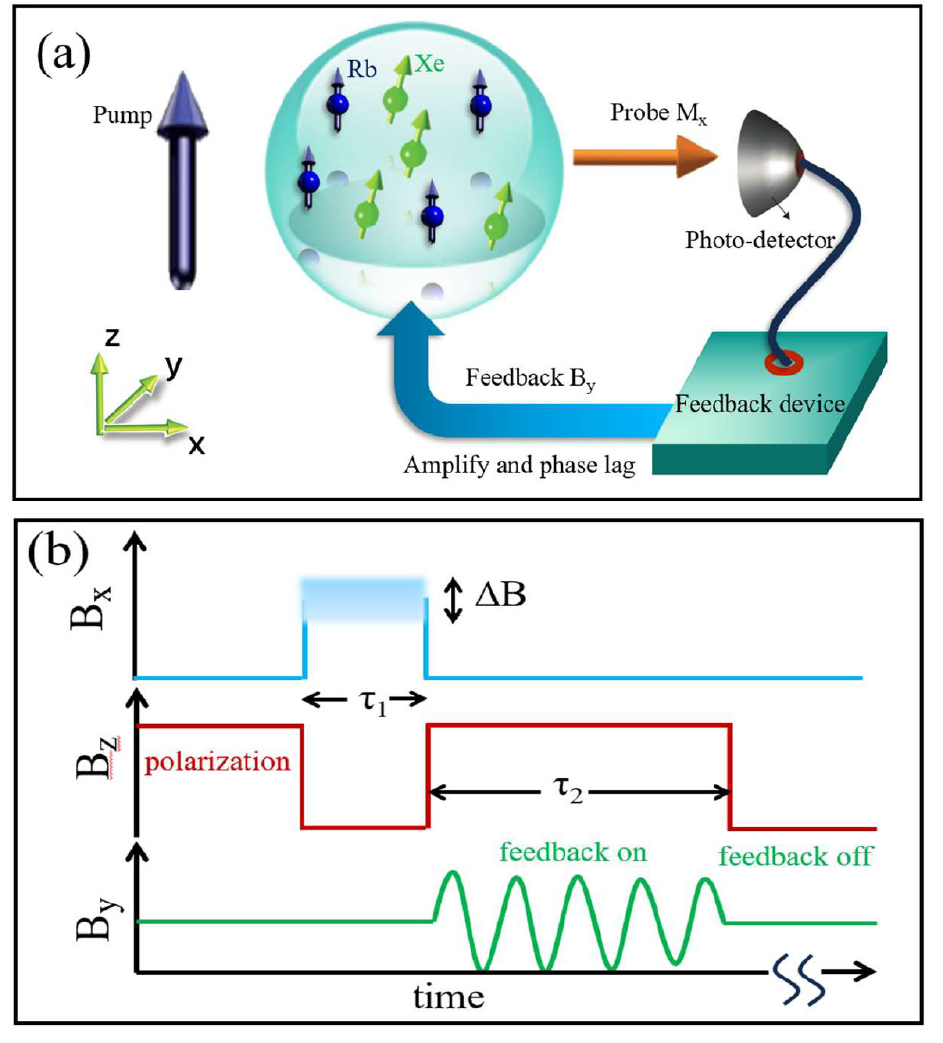}
\caption{Schematic drawing of the experimental setup and time sequence protocol. (a) The vacuum glass cell contains several torrs of $^{129}$Xe and a small droplet of $^{87}$Rb atoms. The wavelengths of the pump and probe lasers are 795 nm and 780 nm, corresponding to the $D_1$ and $D_2$ lines of $^{87}$Rb respectively.   (b) In a single experiment realization, spins are first polarized by optical pumping for 40 $\sim$ 60 seconds. Preliminarily, the benchmarks are set to $I_{x,0}=$ 1000 $\mu$A (current-to-magnetic-field conversion coefficient $C_{\rm B/I}\sim$ 1.55 mG per mA) and $\tau_{1,0}=$ 300 ms, The feedback-off (or depolarization) period lasts for 40 $\sim$ 60 seconds, same  as the polarization period. $\Delta B$ is a perturbation on $B_x$ to test the robustness.}
\label{fig1}
\end{figure}

Motivated by this analogue, in this study, we report an experimental observation of a time crystal phase in a Rb-Xe hybrid spin maser system with a feedback induced retarded interaction. The experimental setup, depicted schematically in Fig.\ref{fig1}, is described in detail in  Methods. Spin maser is a self-driven oscillating atomic system\cite{Bloom1962,Chupp1994PRL,Sato2018PLA}, where a phase coherent feedback is used to maintain the persistent spin oscillation of macroscopic  ensemble of atoms and balance the spin depolarization or decoherence. The spin maser is not only with practical significance in geomagnetic measurements\cite{Dyal1969,Kubo1972} and  magnetic navigation\cite{Canciani2022}, but also of fundamental interest in the searching for permanent electric dipole moment \cite{Harris1999,Chupp2001PRL,Romalis2001} and spin-dependent exotic interactions \cite{2021Floquetmaser}. Unlike the conventional spin maser system where the feedback is used to amplify the signal and maintain spin precession with the Larmor frequency, the feedback in our system with a time delay leads to an effective retarded interaction. This retarded interaction is responsible for a self-sustained oscillation in a macroscopic ensemble with an emergent frequency different from the intrinsic Larmor frequency, but is crucially determined by the phase lag of the time delay feedback. The experimental evidences of the two ingredients of time crystal: its robustness and spontaneous symmetry breaking, have been demonstrated and it is shown that such a time crystal phase only emerges when the feedback strength exceeds a critical value.

\begin{figure}[htbp]
\centering
\includegraphics[width=0.5\textwidth]{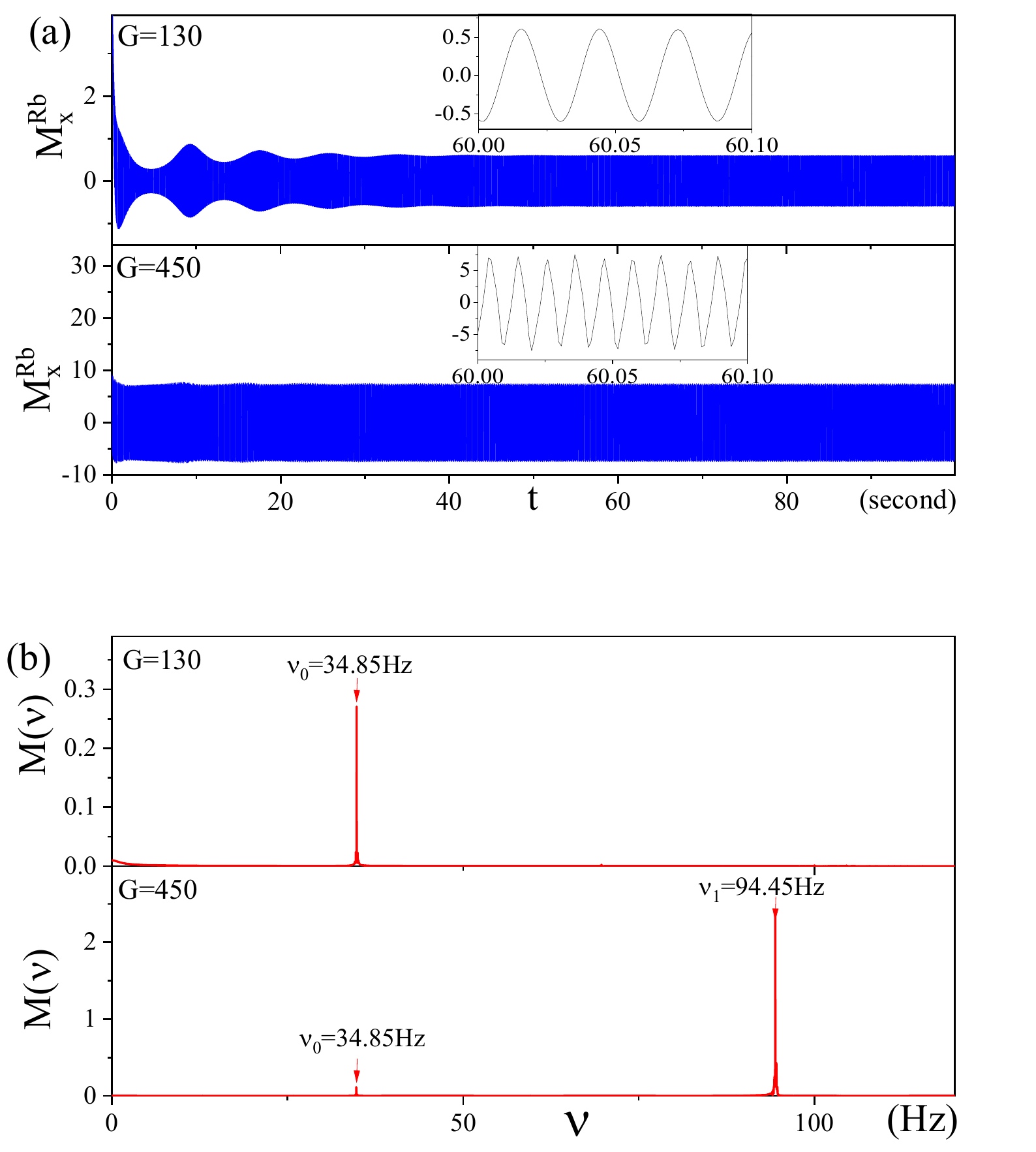}
\caption{Observation of time crystal signal. (a) The dynamics of magnetization of Rb along x-direction ($M_x^{\rm Rb}$) in the presence of weak ($G=130$) and strong ($G=450$) feedback (the insets zoom in on the $t=60$ s vicinity of the axis)  and (b) the corresponding Fourier spectra in frequency domain.  The phase lag is fixed as $\theta=\pi/2$.}
\label{fig2}
\end{figure}

\section{Results and Discussion}
\subsection{Experimental verification of time crystal} 
Fig.\ref{fig2} (a) shows typical real-time signals of spin maser system ($\sim M_x^{\rm Rb}$) with weak and strong feedback strengths respectively. Their corresponding Fourier spectra are plotted in Fig.\ref{fig2} (b).  For a weak feedback ({\it e.g.} $G$ = 130), one can observe a sharp peak at $\nu_0$ in the Fourier spectrum, corresponding to the Larmor frequency of precession around the polarized field $B_z$ ($\nu_0=\gamma_{\rm Xe} B_z\simeq35$ Hz with $B_z \simeq$ 30 mG and $\gamma_\text{Xe} = -2\pi \times 1.178 \ \text{rad} \cdot \text{kHz}\cdot \text{G}^{-1}$ is the magnetogyric ratio for Xe spins\cite{Sato2018PLA, 2021Floquetmaser, 2023PRApplied}).  So in this case,  the feedback is implemented to compensate for the energy loss and the spin depolarization, and sustain the intrinsic Larmor precession of the spin maser system. However, it doesn't induce new signal with other frequency.

In the presence of a strong feedback({\it e.g.} $G$ = 450), however, the situation is different.  In addition to the original peak at $\nu_0$, a stronger signal suddenly emerges at $\nu_1\simeq94$ Hz, which indicates a self-sustained oscillation with an emergent frequency independent of the intrinsic Larmor frequency of the spin maser system, thus provides an evidence of time crystal.  When the feedback is turned off, the $\nu_1$ signal disappears immediately while the $\nu_0$ signal decays exponentially as in the free induction decay mode (non-maser mode).

\begin{figure}[htbp]
\centering
\includegraphics[width=0.5\textwidth]{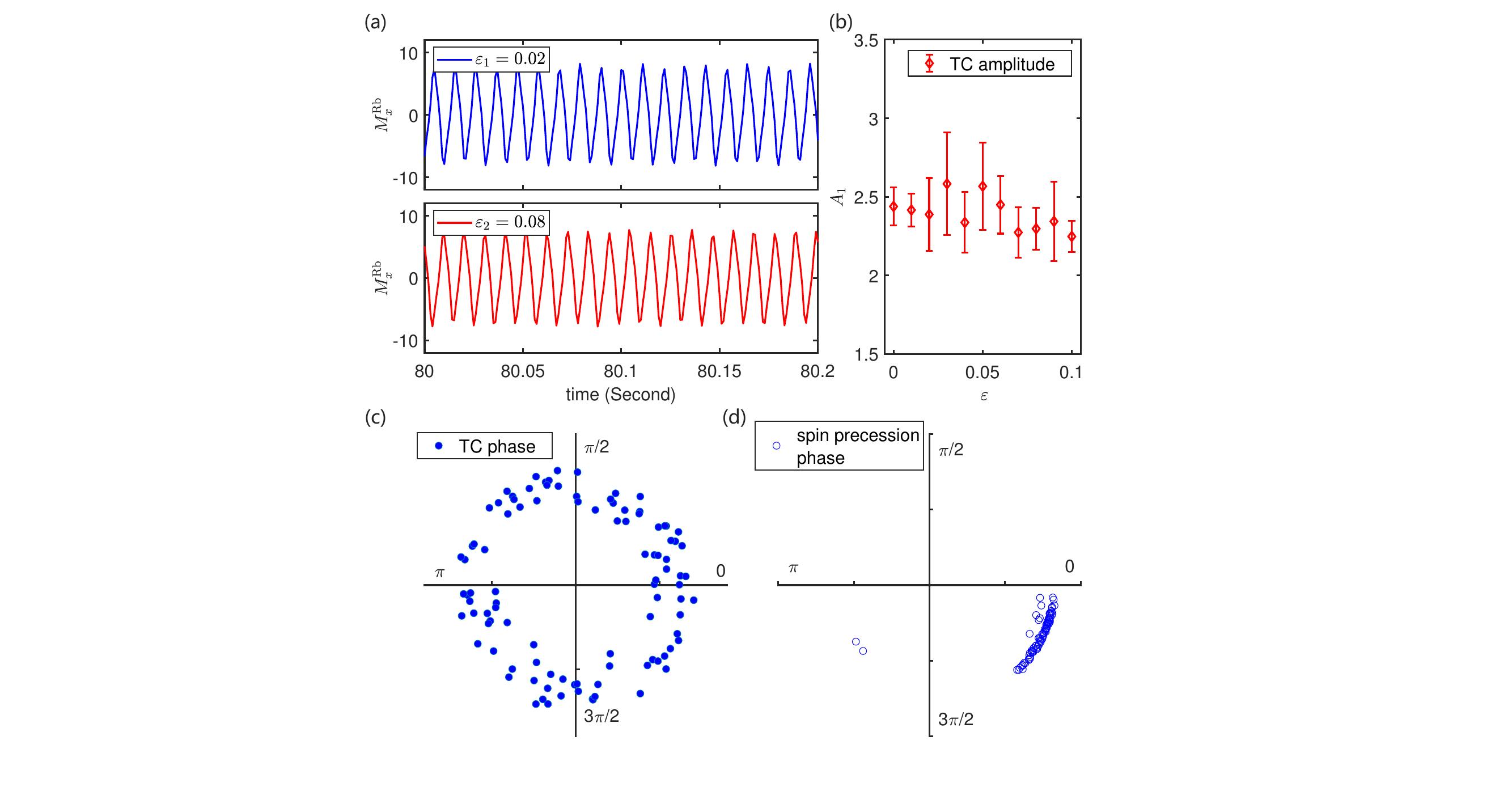}
\caption{Robustness and spontaneous time translation symmetry breaking. (a) The dynamics of $M^{Rb}_x$(magnetization of Rb along x-direction) starting from two different initial states characterized by the perturbation parameter $\varepsilon$. (b) The amplitude of the time crystal (TC) oscillation as a function of $\varepsilon$ characterizing different initial states. Each error bar represents the standard error of 5 repeated measurements under same experiment condition (length of single side equals to the corresponding standard error).(c) and (d) The distribution of the time phases (the axial angles of the solid and empty dots) in 100 physical realizations of (c) the time crystal (TC) oscillation in the case with strong feedback $G=640$ and (d) the intrinsic Larmor precession in the system without feedback ($G$ = 0). The phase lag is fixed as $\theta=\pi/2$.}
\label{fig3}
\end{figure}

One of the ingredient of time crystal is its robustness against perturbation. To demonstrate this point, we exert a perturbation $\Delta B_x$ on in the initialization pulse height $B_x$, thus the rotating angle in the initial state becomes:
\begin{equation}
\alpha=(1-\varepsilon)\gamma B_{x} \tau_1
\label{eq2}
\end{equation}
where the dimensionless parameter $\varepsilon=\Delta B_x/B_x$. As shown in Fig.\ref{fig3} (a) and (b), after the initial transient time, the amplitude of the oscillation for systems with different $\varepsilon$ agree with each other within the error bar (see Fig.\ref{fig3} b), while their time phases are different.

The time crystal is also associated with a spontaneous time translation symmetry breaking, which means that  the time  phases of the oscillations in independent experimental realizations take random values with an uniform distribution between $0$ and $2\pi$. In contrast, the relative time phases in oscillations induced by polarized magnetic field ($B_z$) takes a fixed value depending on the initial polarization. To verify this point experimentally, we fix $G=640$ and $\theta=\pi/2$ in the strong feedback regime, repeat the experimental realization for $\mathcal{N}$ times ($\mathcal{N}=100$), and extract the relative phase from the oscillation in each experimental realization. These relative phases are plotted as the axial angle in Fig.\ref{fig3} (c), which exhibits a random distribution between $0$ and $2\pi$, agreeing with the criteria of spontaneous symmetry breaking. As a comparison, we also measure the relative phases without feedback ($G=0$), whose distribution is within a narrow regime as depicted in Fig.\ref{fig3} (d), indicating the absence of spontaneous symmetry breaking in this case.

\begin{figure}[htbp]
\centering
\includegraphics[width=0.48\textwidth]{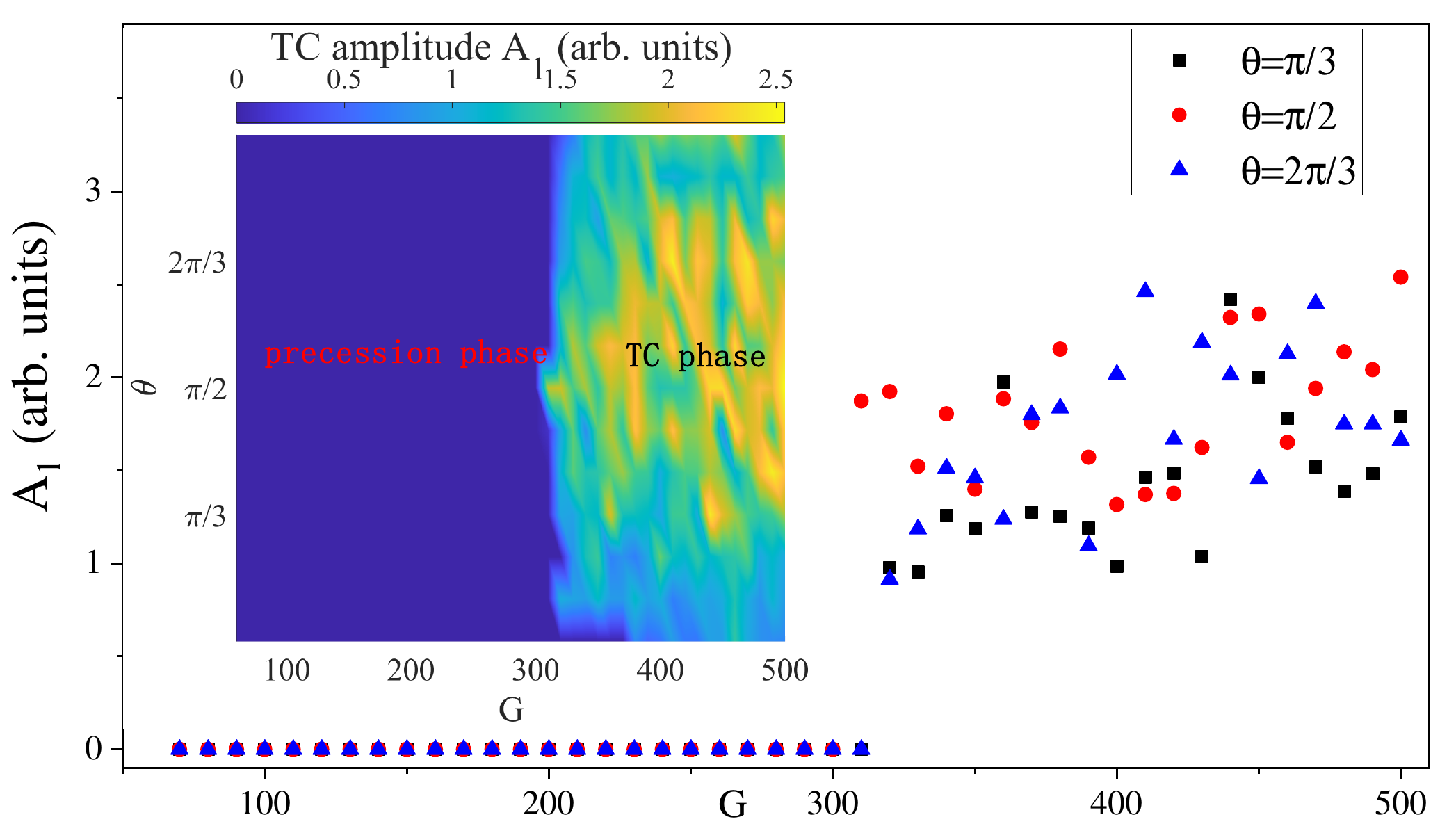}
\caption{Time crystal phase transition and phase diagram. The amplitude of the time crystal (TC) oscillation $A_1$ as a function of $G$ (feedback strength factor) in the presence of different $\theta$ (the phase lag between the feedback field $B^f_y$ and the measured polarization $M^{\rm Rb}_x$), a 1st order phase transition occurs at $G\simeq$ 300. The inset indicates the amplitude as a function of $G$  and $\theta$ and the global phase diagram.}
\label{fig4}
\end{figure}

\subsection{Dynamical phase transition and the phase diagram}
After comparing the different dynamical phases in weak and strong feedback regime, we now focus on the phase transition between them as well as the phase diagram in terms of the two control parameters of our feedback system, the feedback strength $G$ and the phase lag $\theta$. Since the presence of time crystal is companied by emergence of the peak at $\nu_1$, we choose the height of the $\nu_1$ peak in the fourier spectrum $A_1$ as an order parameter to characterize the time crystal phase. For different $\theta$, $A_1$ as a function of $G$ is plotted in Fig.\ref{fig4}, from which we can find a sudden jump from zero to a finite value occurs at a critical value of $G=G_c$, indicating that the system experiences a discontinuous phase transition. We notice that at the critical point, the corresponding feedback magnetic field $B_f^y\simeq B_z$, indicating the the time crystal emerges when the feedback field surpasses the original magnetic field. The phase diagram in terms of $G$ and $\theta$ is also plotted in the inset of Fig.\ref{fig4}, which indicates that in a large regime of $\theta$ ($\pi/3<\theta<5\pi/6$), the corresponding $G_c$ barely depends on $\theta$.

\begin{figure}[htbp]
\includegraphics[width=0.48\textwidth]{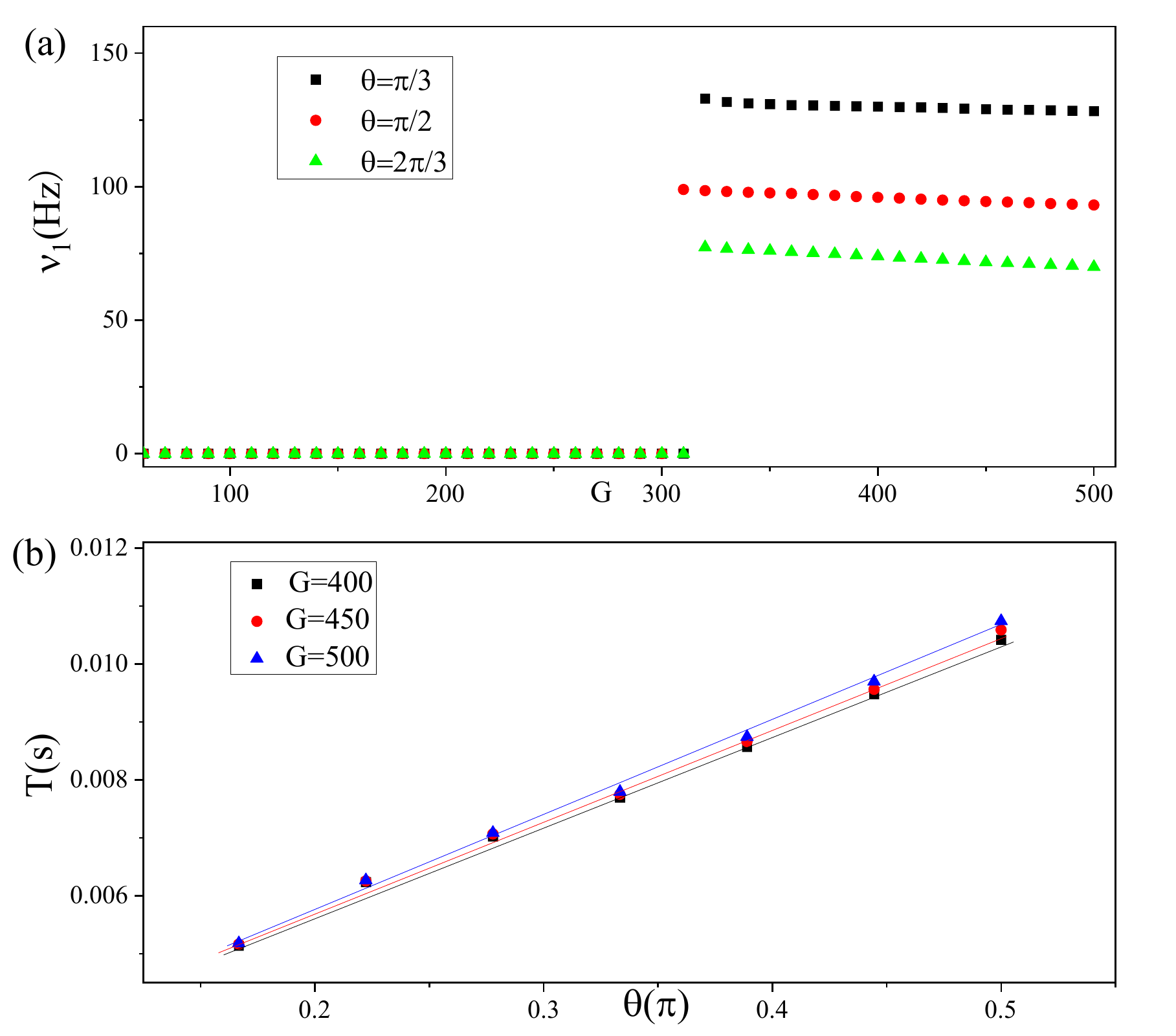}
\caption{Dependence of time crystal features on the control parameters. (a) The frequency of the time crystal oscillation ($\nu_1$) as a function of feedback strength $G$ in the presence of different phase lag $\theta$. (b) The period of time crystal $T=1/\nu_1$ as a function of the phase lag $\theta$ for different $G$. }
\label{fig5}
\end{figure}

Besides the amplitude, the frequency of the emergent self-sustained oscillation ($\nu_1$) is another important feature to characterize time crystal, and we will study the dependence of $\nu_1$ on the control parameters in our feedback system.  $\nu_1$ as a function of $G$ for different $\theta$ is plotted in Fig.\ref{fig5} (a), from which we can see that once the system enter the time crystal phase, $\nu_1$ barely depends on $G$, but is crucially determined by $\theta$.

As we stated above, the feedback procedure in our setup leads to an effective retarded interaction with a characteristic time scale (the delay time $\tau$), which is responsible for the emergent period of the time crystal phase.  To figure out the relationship between the delay time (proportional to the phase lag $\theta$ in Eq.(\ref{eq3}) and the time crystal period $T=2\pi/\nu_1$, we plot $T$ as a function of $\theta$ for different $G$ in Fig.\ref{fig5} (b), which approximately exhibit linear relations, whose slopes slightly depends on the value of $G$. This result shows that the ``lattice constant'' of the time crystal is indeed proportional to  the characteristic time scale in the feedback-induced retarded interaction.

\subsection{Discussion}
Finally, we add some remarks about the spontaneous time translation symmetry breaking, which is usually referred to persistent oscillations emerging from systems without explicit time-dependence. Even though our system is ``time dependent'', the time dependence
enters our system through the relative time difference (delay time) in terms of two-body interactions which are still translational invariant, in contrast to the Floquet engineering  whose time-dependence is explicitly through overall time. As a consequence, such a  retarded interaction induced time crystal is closer to the original idea of crystal, compared to  those time crystal mechanisms based on nonlinearity, which comes from either a mean-field \cite{Collado2021} or a coarse-grained treatment\cite{Gong2018,Yang2023} of the interaction. Another evidence of the spontaneous symmetry breaking in our system is that the time crystal phase emerges only when the feedback strength exceeds a critical value, a reminiscence of certain  solid state systems where the charge density wave order can only emerges for a sufficiently strong interaction.

\section{Conclusion} 
In conclusion, in a Rb-Xe hybrid spin maser system with time-delay feedback,  we observed a new type of time crystal, which straightforwardly generalizes the concept of crystal into time domain. This result demonstrates that the feedback, besides being a signal amplifier and stabilizer, can also lead to intriguing non-equilibrium phases of matter when the retardation effect is included. Apart from its fundamental interest, the potential technological application of this retardation-induced time crystal phase in time metrology and precision measurement can also be envisioned.

\section{Methods}
To observe the time crystal in a spin maser system, we prepare a thermal ensemble of Rb-$^{129}$Xe hybrid atomic spins, as illustrated in Fig.\ref{fig1}(a). As the first step,  a hot ($\sim$120 $^{\circ}$C) gaseous Rb-Xe vapor (containing a droplet of natural abundance Rb, 5 torr of isotope enriched $^{129}$Xe and 50 torr of buffer gas $\rm N_2$) is pumped with a resonant laser along the $z$ axis, Rb spins are first polarized and subsequently, this polarization is transferred  to Xe spins via rapid spin-exchange collision. Secondly, the Xe spins precess around a dc magnetic field $B_z$ along the $z$ axis. The precession is followed by Rb spins and read out by an off-resonant laser with an optical polarimeter protocol in the $x$ axis \cite{Budker2002RMP}. At last, the photodetector signal is processed by an electric circuit, which can amplify the ac signal with a tunable phase lag. After that, the signal is fed back via a magnetic coil along the $y$ axis, which, in turn drives the Xe spins coherently.

For a single experimental realization as depicted in Fig.\ref{fig1}(b), we prepare the initial state by first utilizing 795-nm pump laser in the $z$ axis with power of approximately 40 mW, which polarizes  the spins in the system along z-direction,  then imposing  a magnetic pulse along x-direction ($B_x$) with a short duration $\tau_1$ on the ensemble to rotate the spin around the x-direction by an angle $\alpha =\gamma B_x \tau_1$, with $\gamma$ being the magnetogyric ratio.

After  preparation of such a polarized state, we apply a magnetic field  along the $z$ axis ($B_z$) with a duration $\tau_2 \gg \tau_1$,  and at the same time, continuously monitor the polarization along x-direction of the Rb atomic spins. This signal is transmitted to the feedback device, where the  original signal from the vapor cell is processed then fed back to the system parameter via the $B_y$ coil. Such a feedback system
is devised to produce a transverse magnetic field $\mathbf{B}_f = (0, B^f_y, 0)$ in response to a measured polarization $M^{Rb}_x$ as:
\begin{equation}
B^f_y(t)=\frac{C_y G}{R} V_{pd}(t-\frac{\theta}{\nu_0}) \label{eq3}
\end{equation}
where $R$ is the impedance of load (including a resistor and an inductive magnetic coil), $G$  represents the electric circuit gain factor and $C_y$ ($\sim$ 2 mG per mA)  is the coil coefficient of the $B_y$ field. $V_{pd}$ is photodetector output voltage, which is proportional to  $M^{\rm Rb}_x$.  $\nu_0$ is the intrinsic Larmor frequency of the spin maser system which is proportional to $B_z$. $\theta$ is the phase lag between the feedback field $B^f_y$ and the measured polarization $M^{\rm Rb}_x$, which is proportional to the delay time in the feedback process.  The parameters in our feedback system is highly tunable: the amplifier gain factor $G$ ranges from 0 to 640 and phase lag $\theta$ is from $\pi/9$ to $5\pi/6$. The probe laser power is kept down to $\simeq$ 1 mW to prevent signal saturation in the photodetector.

 In the final step of the experiment realization, the feedback and the polarized magnetic field ($B_z$) are turned off,  thus the system is depolarized and return to the initial state, from which we can start over again to perform another experimental realization.

\section{Data Availability} 
The data and code supporting the figures in this study are available upon reasonable request from the corresponding author.

\section{Acknowledgement} 
GL acknowledge the support by Chinese Academy of Sciences (Grant No. E209YC1101). ZC is supported by the National Key Research and Development Program of China (Grant No. 2020YFA0309000), NSFC of  China (Grant No.12174251), Natural Science Foundation of Shanghai (Grant No.22ZR142830),  Shanghai Municipal Science and Technology Major Project (Grant No.2019SHZDZX01)

\section{Author contributions}
Weiyu Wang, Qianjin Ma and Erwei Li did the experiment. Weiyu Wang, Mingjun Feng, Zi Cai and Guobin Liu wrote the paper. Guobin Liu conceived the idea and supervised the project. 
\section{Ethics declarations}
This research does not contain any content on human or animal.
\subsection{Competing interests}
The authors declare no competing interests.

\end{document}